\documentclass{iaus}
\usepackage{graphicx}

\newcommand{\apj}{{\it ApJ}}
\newcommand{\apjs}{{\it ApJS}}
\newcommand{\aj}{{\it AJ}}
\newcommand{\mnras}{{\it MNRAS}}
\newcommand{\aanda}{{\it A\&A}}

\title[IAUS266.~~Star clusters with dual red clumps] 
{Star clusters with dual red clumps}

\author[L. Girardi, S. Rubele \& L. Kerber]   
{L\'eo Girardi,$^1$ Stefano Rubele$^{1,2}$ \and Leandro Kerber$^3$}

\affiliation{$^1$Osservatorio Astronomico di Padova, Italy \\
$^2$Dipartimento di Astronomia, Universit\`a di Padova, Italy \\ 
$^3$Universidade Estadual de Santa Cruz, Ilh\'eus, Brazil}

\pubyear{2010}
\volume{266}  
\pagerange{1--6}
\jname{Star clusters: basic galactic building blocks}
\editors{R. de Grijs \& J. R. D. L\'epine, eds.}
\begin{document}

\maketitle

\begin{abstract}
A few star clusters in the Magellanic Clouds exhibit composite
structures in the red-clump region of their colour--magnitude
diagrams. The most striking case is NGC~419 in the Small Magellanic
Cloud (SMC), where the red clump is composed of a main blob as well as
a distinct secondary feature. This structure is demonstrated to be
real and corresponds to the simultaneous presence of stars which
passed through electron degeneracy after central-hydrogen exhaustion
and those that did not. This rare occurrence in a single cluster
allows us to set stringent constraints on its age and on the
efficiency of convective-core overshooting during main-sequence
evolution. We present a more detailed analysis of NGC~419, together
with a first look at other populous Large Magellanic Cloud clusters
which are apparently in the same phase: NGC~1751, NGC~1783, NGC~1806,
NGC~1846, NGC~1852 and NGC~1917. We also compare these Magellanic
Cloud cases with their Galactic counterparts, NGC~752 and NGC~7789. We
emphasise the extraordinary potential of these clusters as {\em
absolute} calibration marks on the age scale of stellar populations.
\end{abstract}

\firstsection 
\section{NGC~419 and its dual red clump}

\begin{figure}[b]
\begin{center}
 \includegraphics[width=0.8\textwidth]{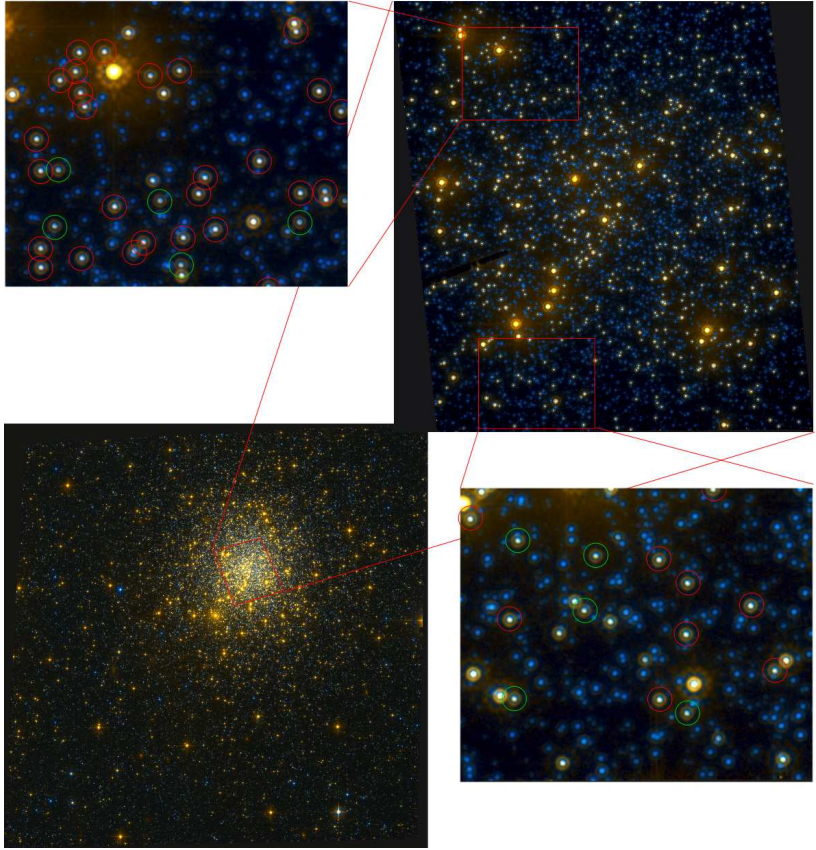} 
 \caption{False-colour images (in the electronic version) of the Small
 Magellanic Cloud star cluster NGC~419, derived from ACS/WFC {\it
 (bottom left)} and HRC {\it (top right)} images in the F555W and
 F814W filters. The top left and bottom right panels zoom in onto the
 HRC image. Red-clump stars are marked with circles. At first glance,
 it is evident that they are quite homogeneous in their colours and
 luminosities, as expected for red-clump stars. However, when
 comparing their first Airy rings, one notices some quite subtle and
 {\em systematic} differences in their brightnesses. Indeed, there are
 two kinds of red-clump stars: the most numerous and brighter, and a
 subsample (about 15\% of the total) of fainter `secondary red-clump
 stars' (marked with red and green circles, respectively, in the
 electronic version). The luminosity difference between these groups
 is about 0.4~mag. There is also a $\sim0.04$~mag difference in the
 mean colour which, however, is too small to be noticeable in the
 figure. The two kinds of red-clump stars are uniformly distributed
 across the HRC image. } \label{fig_image}
\end{center}
\end{figure}

NGC~419 is a populous star cluster located to the east of the Small
Magellanic Cloud (SMC)'s bar in a region relatively devoid of dust and
free from contamination by the SMC field. There are two wonderful
pairs of {\sl HST} images of this cluster, taken in the F555W and
F815W filters, originally obtained as part of programme GO-10396 (PI
J. S. Gallagher) and now retrievable from the {\sl HST} archive. They
are shown in Figure~\ref{fig_image}.  While the ACS/WFC images reveal
the overall cluster structure, the ACS/HRC observations show the
details of the cluster core.

\begin{figure}[b]
\begin{center}
 \includegraphics[width=0.6\textwidth]{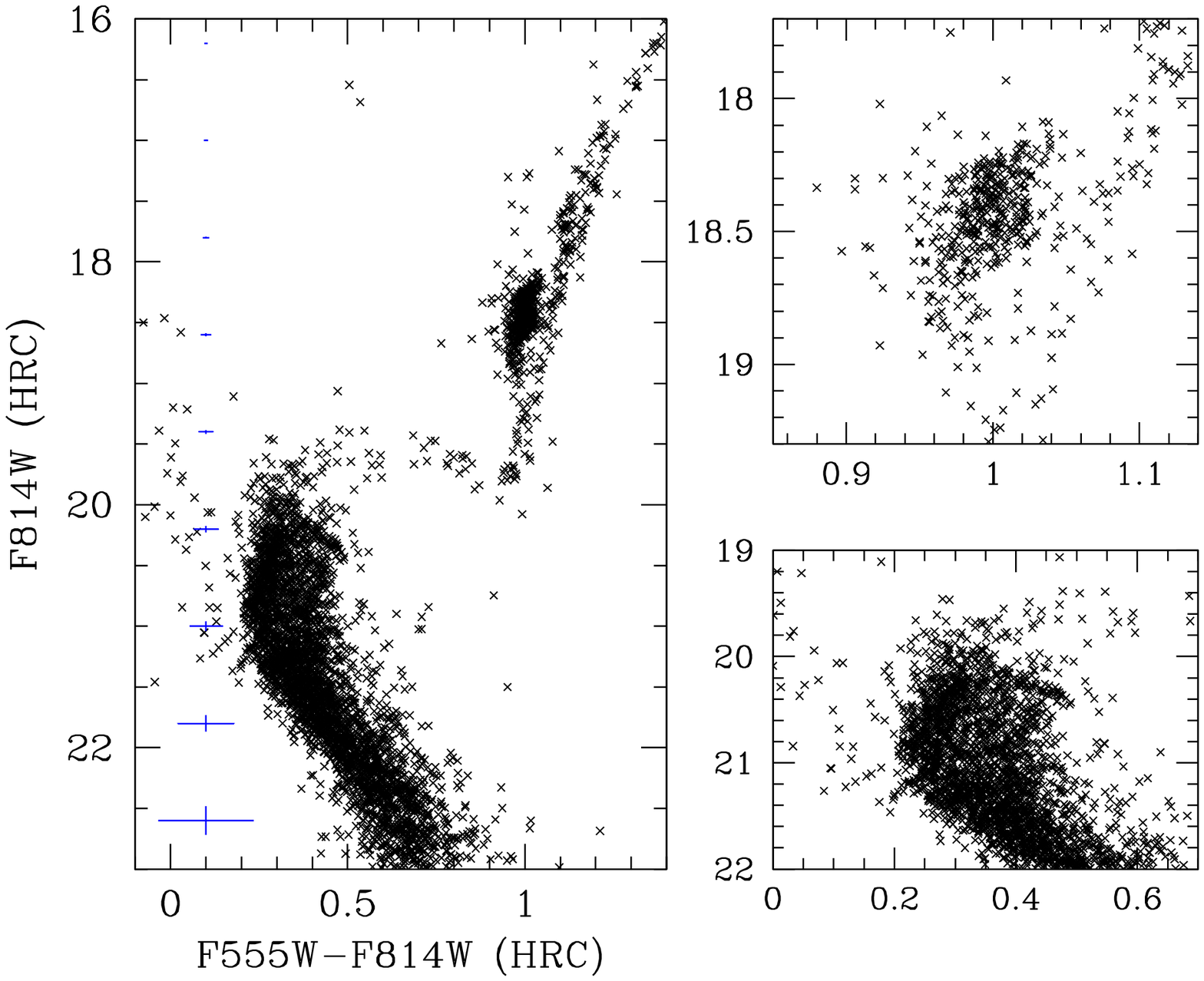} 
 \caption{NGC~419 colour--magnitude diagram (CMD; {\it left panel}),
 derived from the ACS/HRC data (Girardi et al. 2009). The error bars
 in the left panel are upper limits to the photometric errors. The two
 right-hand panels detail the red-clump and main-sequence turnoff,
 clearly showing their composite structure.}
 \label{fig_cmd}
\end{center}
\end{figure}

The HRC images allow us to perform photometry of an impressive
quality, and reveal at least two surprises: the presence of multiple
main sequence turnoffs (MMSTOs), and a composite red clump which
contains a pronounced faint extension (Figure~\ref{fig_cmd}). Both
features were noticed by Glatt et al. (2008). They tentativeley
interpreted the faint red-clump structure as being caused by SMC field
contamination. However, this explanation does not stand up to a simple
star count in the neighbouring field selected from the ACS/WFC image
(see Girardi et al. 2009): the expected number of field red-clump
stars in the ACS/HRC area is just 4.5, while the faint red clump
contains at least 50 objects (Figure~\ref{fig_cmd}). Assuming a
Poissonian distribution for the field stars, the probability ($P$)
that these 50 stars are drawn from the SMC field alone is virtually
zero ($P<10^{-9}$).

Detached binaries cannot give rise to this faint red clump either,
since any combination of single stars will cause an extension of the
red clump to brighter magnitudes.

What, then, is the dual red clump of NGC~419 made of? Girardi et
al. (2009) show that it is caused by the presence of stars following
two very different evolutionary paths. In the following, we will
expand their arguments a little. There are, in fact, a few different
ways of looking at dual red clumps, as detailed below.

\section{Electron degeneracy on the scene}

The basic fact behind dual red clumps is the dilemma facing low- to
intermediate-mass stars after exhausting their central hydrogen and
departing from the main sequence: {\em to RGB or not to RGB, that is
the question!} (RGB: red-giant branch; adapted from Shakespeare
1600). Stars slightly more massive than 2~M$_\odot$ simply contract
their hydrogen-exhausted cores until a temperature of about $10^8$~K
is reached. This causes core-helium ignition and the overall expansion
of the star, which quickly settles in the red clump and quietly burns
helium thereafter. Stars slightly less massive instead suffer from
electron degeneracy in the core before helium can be ignited. Electron
pressure stops core contraction and the core becomes nearly
isothermal. Subsequently, neutrino losses cool its very centre. The
result is that He ignition is prevented for a long time. These stars
have to climb all the way up to the RGB tip, where helium ignites in a
flash, finally freeing their cores from electron degeneracy.

The difference between these two evolutionary paths is a quite abrupt
function of initial mass (see Sweigart et al. 1990; Girardi 1999). In
an interval of order 0.2~M$_\odot$, the core mass at He ignition
changes from 0.46 to 0.33~M$_\odot$, depending on the presence or
absence of electron degeneracy. This core-mass difference alone causes
a $\sim0.4$~mag difference in the initial luminosity of He-burning
stars (with the red clump of the younger population being fainter),
which is adequate to explain the two clumps of NGC~419.

The quantitative details of this explanation, based on stellar tracks
and isochrones including the complete He-burning phase and computed
with a fine mass resolution, can be found in Girardi et al. (2009; see
also Girardi 1999).

\section{Just another manifestation of prolonged star formation}

From the discussion above, one may already obtain a hint that dual red
clumps are a common feature in galaxy fields containing stars of all
initial masses (ages), provided that such fine CMD features are not
blurred by effects such as differential reddening, photometric errors
and metallicity dispersions (Girardi 1999). Indeed, striking examples
of faint secondary clumps are provided in the {\sl Hipparcos}
solar-neighbourhood CMD (Girardi et al. 1998) and by the CMDs of some
outer Large Magellanic Cloud (LMC) fields (Bica et al. 1998; Piatti et
al. 1998).

On the other hand, it is also clear that dual red clumps were {\em not
expected} to be found in single star clusters, since for a single age
the dispersion in turnoff masses is negligible and much smaller than
the $\sim0.2$~M$_\odot$ required to explain the presence of dual red
clumps. This was concluded by Girardi et al. (2000) in their study of
the open clusters with dual red clumps, NGC~752 and NGC~7789.

However, we now have convincing evidence that star clusters are not
single-aged stellar populations. In addition to the old globular
clusters (see Piotto, this volume) and NGC~419 itself, many
intermediate-age clusters in the LMC show MMSTOs. The simplest
explanation of MMSTOs is based on prolonged periods of star formation
(Bertelli et al. 2003; Baume et al. 2007; Mackey et al. 2008; Milone
et al. 2009).

Bastian \& de Mink (2009) recently provided an alternative explanation
for the presence of MMSTOs, based on the $T_{\rm eff}$/colour spread
of main-sequence stars with different rotation rates. The idea is, no
doubt, very interesting and worth of further exploration. One crucial
point to be clarified is whether rotation may cause the small spread
in core mass at the end of the main sequence that is {\em required} to
explain the dual red clump in NGC~419. In other words, the different
rotation velocities on the main sequence have to produce a direct
effect in the mass of H-exhausted cores, otherwise Bastian \& de
Mink's (2009) explanation cannot be applied to NGC~419.
 
\begin{figure}[b]
\begin{center}
 \includegraphics[width=0.5\textwidth]{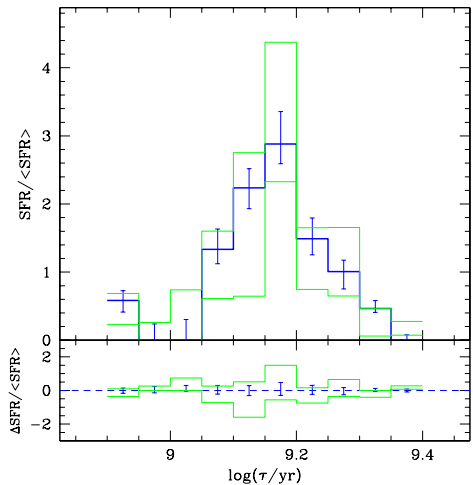} 
 \caption{Star-formation history of NGC~419, as determined based on
 the classical method of CMD reconstruction (Rubele et al., in
 prep.). The heavy line with error bars is the best solution with its
 68\% confidence levels, reflecting random errors only. The top and
 bottom histograms instead indicate the total systematic errors
 derived from the full range of possible distances, reddenings and
 metallicities of the cluster.} \label{fig_sfh}
\end{center}
\end{figure}

Rubele et al. (in prep.) go a little further into this question. They
apply to NGC~419 the classical method of star-formation-history
recovery through CMD reconstruction (cf. Kerber et al. 2009), deriving
a star-formation rate which lasts for at least 500~Myr
(Figure~\ref{fig_sfh}). The spread in turnoff masses turns out to be
as large as 0.18~M$_\odot$. A $\chi^2$-like statistic reveals that
their best-fitting solutions are {\em excellent} representations of
the data. It remains to be clarified whether the rotation advocated by
Bastian \& de Mink (2009) will produce similarly good descriptions of
the CMDs of clusters with MMSTOs, keeping at the same time the number
of free parameters in the models at an acceptable level.

\section{A new way of weighing stellar cores}

The above interpretation implies that stars leaving the NGC~419 main
sequence have an H-exhausted core of almost exactly
0.33~M$_\odot$. The uncertainty in this quantity is as small as
$\sim0.01$~M$_\odot$. It only depends on basic principles of stellar
structure (e.g., Sch\"onberg \& Chandrasekhar 1942) and on the
well-known physics of partially degenerate matter. One may consider
NGC~419 observations as providing the most precise (although indirect)
determinations of the H-exhausted core mass of living stars.

Asteroseismology will eventually provide direct measurements of the
core mass of individual red giants in the solar vicinity (see, e.g.,
Dupret et al. 2009), but at the cost of considerable investments in
terms of dedicated telescopes and data analysis. The NGC~419 results,
in contrast, come almost for free.

\section{An absolute mark on the age scale of stellar populations}

A direct consequence of knowing the core mass of a star at the end of
its main-sequence evolution, is that one has a way of measuring the
extent of convective-core overshooting, a phenomenum that has plagued
age determinations of star clusters for decades. One just has to
combine the red-clump observations (indicative of 0.33~M$_\odot$
H-exhausted cores) with some indicator of the main-sequence mass,
which can be, very simply, the main-sequence location in the CMD. And
once convective-core overshooting is constrained, age determinations
become much more solid.

Girardi et al. (2009) pursued this idea by simultaneously fitting the
red-clump morphology and the colour difference between the
main-sequence termination and the red clump in NGC~419. The best
simultaneous fit was found for a cluster mean age of
$t=1.35_{-0.04}^{+0.11}$~Gyr and an efficiency of convective
overshooting of $\Lambda_{\rm c}\!=\!0.47_{-0.04}^{+0.14}$ pressure
scale heights (see Bressan et al. 1993 for the formalism). It is worth
mentioning that our results allow us to exclude the classical
Schwarzschild criterion for convective boundaries with very high
confidence.

It is remarkable that the above-mentioned age determination is quite
independent of distance, reddening, binary fraction {\em and}
overshooting (see Girardi et al. 2009). Hence, star clusters with dual
red clumps have great potential of becoming {\em absolute marks} on
the age scale of stellar populations.

We recall that absolute age marks are extremely rare in stellar
astrophysics. Accurate absolute ages can be derived for young
associations on the basis of their kinematics (for ages up to $\sim4$
Myr; e.g., Brown et al. 1997), and for old globular clusters once
their upper age limit is constrained by cosmological measurements
(e.g., $<13.73$~Gyr; Spergel et al. 2007). Inside this very wide age
interval, ages of stellar aggregates depend somewhat on the extension
of convective boundaries, which are notoriously uncertain from a
theoretical point of view. Clusters with dual red clumps may represent
the exception to this rule, providing absolute marks at ages of about
1.3~Gyr.

\section{NGC~419 is not alone}

The analysis in Girardi et al. (2009) is about to be repeated with
up-to-date stellar models (Bressan et al., in prep.), and removing the
assumption of a constant star-formation rate during the period of
formation of the MMSTOs. The error bars in the age determination will
inevitably increase compared to Girardi et al. (2009). Moreover, we
have recently realised that the uncertainty in the metal content
represents a good fraction of the error budget for NGC~419 (Rubele et
al., in prep.). More clusters with dual red clumps, with as far as
possible good [Fe/H] determinations, are necessary to obtain tighter
constraints on their ages and overshooting efficiency.

Such star clusters exist and can even be numerous in the Magellanic
Clouds. A quick look at the {\sl HST}/ACS CMDs from Milone et
al. (2009) reveals that dual red clumps are present in the LMC star
clusters NGC~1751, NGC~1783, NGC~1806, NGC~1846, NGC~1852 and NGC~1917
(Girardi et al. 2009). All exhibit MMSTOs at a magnitude which is
compatible with the lower limit for the settling of electron
degeneracy. We are now examining their archival {\sl HST} data,
checking accurately how much of the faint clump features can be
attributed to contamination from the LMC field, and then repeating the
same analysis as for NGC~419.

In our Galaxy, the open clusters NGC~752 and NGC~7789 contain, among
their radial-velocity members, a handful of stars which apparently
belong to a faint secondary clump (Mermilliod et al. 1998; Girardi et
al. 2000). Other candidates are NGC~2660, NGC~2204 (Girardi et
al. 2000) and Tr~20 (Seleznev et al. 2009). In all these cases, the
small numbers of red-clump stars may hamper the derivation of clear
constraints to the stellar models and cluster ages. The situation with
the populous Magellanic Cloud clusters is clearly different, and much
more exciting.

\begin{acknowledgements}
The {\sl HST} data illustrated in this talk were obtained from the
Multimission Archive at the Space Telescope Science Institute
(STScI). STScI is operated by the Association of Universities for
Research in Astronomy, Inc., under NASA contract NAS5-26555. We thank
the IAU 266 Scientific Organising Committee for the invited talk slot,
I. Platais, P. Marigo, J. D. do Nascimento, G. Carraro and A. Miglio
for their comments and acknowledge the support from INAF/PRIN07 CRA
1.06.10.03, contract ASI-INAF I/016/07/0, and the Brazilian agencies
CNPq and FAPESP.
\end{acknowledgements}

\end{document}